\documentclass[12pt]{article}
\title{Einstein and Bell, von Mises and Kolmogorov: reality and locality,
frequency and probability}

\author{Andrei Khrennikov \\International Center for Mathematical
Modeling \\ in Physics and Cognitive Sciences,\\
University of V\"axj\"o, S-35195, Sweden\\
Email:Andrei.Khrennikov@msi.vxu.se}

\date{}
\begin{document}
\maketitle

\bigskip

{\small We perform frequency analysis of the EPR-Bell argumentation.
One of the main consequences of our investigation is that
the existence of probability distributions of the Kolmogorov-type which was supposed by some
authors is a mathematical assumption which may not be supported by actual
physical quantum processes. In fact, frequencies for hidden variables
for quantum particles and measurement devices may fluctuate
from run to run of an experiment. These fluctuations of frequencies for 
micro-parameters
need not contradict to the stabilization
of frequencies for physical observables. If, nevertheless, micro-parameters are
also statistically stable,
then violations of Bell's inequality and its generalizations
 may be a consequence of dependence of collectives corresponding to two different measurement
devices. Such a dependence implies the violation of the factorization rule for
the simultaneous probability distribution. Formally this rule coincides with the well known
BCHS locality condition (or outcome independence condition). However, the frequency
approach implies totally different interpretation of dependence. It is not dependence
of events, but it is dependence of collectives. Such a dependence may be induced by 
the same preparation procedure.}

\section{INTRODUCTION}

The theoretical and experimental disagreement between Bell's inequality and its generalizations, see,
for example, [1]-[3], and the quantum-mechanical predictions for correlation function has been the origin of
much dispute and speculation. Among the possible explanations for this disagreement the following well-known
ones may be mentioned: 1) impossibility to use {\it local realism}, [1]-[3]; 2) use of {\it probabilistic
assumptions} which may be not supported by actual quantum processes, [4]-[8]. In this paper we continue
to study possible probabilistic sources of the mentioned disagreement. We perform frequency analysis
of the EPR-Bell argumentation. One of the main consequences of this analysis is that
the existence, see [1]-[3], of probability distributions of the Kolmogorov-type [9] 
is a mathematical assumption which may not be supported by actual
physical quantum processes.

The frequency approach in Bell's framework was used in many papers, see,
for example, Stapp, Eberhard, Peres in [2]. In fact, in the most works on Bell's inequality probabilities
are finally identified with frequencies, simply in order to be able to compare with the experimental data.
The main distinguishing feature of our frequency analysis is the study of frequency behaviour
not only on the level of physical observables, but also on the level of hidden variables.
It seems that such a frequency investigation has not been performed. It should be also remarked
that we use the well developed frequency formalism of R.von Mises [10]. This formalism
is not reduced to the frequency definition of probability. In Bell's framework we study such delicate problems
as {\it combining} of collectives corresponding to different measurement devices and 
difference between the frequency and conventional viewpoints to {\it independence.} Different viewpoints to
independence induce different interpretations of Bell-Clauser-Horne-Shimony (BCHS) `locality
condition' [1]- [3], namely the factorization condition 
\begin{equation}
\label{f0}
{\bf p}(A= \epsilon_1,\lambda=k,  B=\epsilon_2)= {\bf p}(A= \epsilon_1,\lambda=k)\; {\bf p}(B= \epsilon_2,\lambda=k)\;,
\end{equation}
where $A$ and $B$ are physical observables corresponding to
two settings of two measurement apparatuses in the EPR-Bohm framework; here $\epsilon_j= \pm 1$
are measurement outcomes on spin-1/2 systems.
In the frequency approach independence
is not independence of events, but independence of collectives. Hence, a violation of the BCHS
factorization condition could not be interpreted as dependence of events corresponding to
measurements for two spatially separated particles. Such a violation is a consequence of dependence
of collectives corresponding to correlated particles.

\section{FREQUENCY PROBABILITY THEORY}

The frequency definition of probability is more or less standard in quantum theory;
especially in the approach based on preparation and measurement procedures, [3], [4]. 
For instance, we can refer to Peres' book [3]:
{\small "If we repeat the same preparation procedure many times, the {\it probability} of a given outcome
is its {\it relative frequency,} namely the limit of the ratio of the number of occurrences of that outcome to the
total number of trials, when these numbers tend to infinity. This ratio {\it must}
tend to a limit if we repeat the same preparation."}
This section contains an introduction to frequency probability theory, see [10] for
the details.

 Let us consider a sequence of physical systems
$
\pi= (\pi_1,\pi_2,..., \pi_N,... )\;.
$
Suppose that elements of $\pi$ have some property, for example, position, 
and this property
can be described by  natural numbers: $ L=\{1,2,...,m \},$ 
the set of labels.  Thus, for each $\pi_j\in \pi,$ we have a number $x_j\in L.$
So $\pi$ induces a sequence 
\begin{equation}
\label{la1}
x=(x_1,x_2,..., x_N,...), \; \; x_j \in L.
\end{equation}
For each fixed $\alpha \in L,$ we have the relative frequency
$\nu_N(\alpha)= n_N(\alpha)/N$ of the appearance of $\alpha$
in $(x_1,x_2,..., x_N).$ Here $n_N(\alpha)$ is the number of elements in
$(x_1,x_2,..., x_N)$ with $x_j=\alpha.$
R. von Mises said that
$x$ satisfies to the principle of the {\it statistical stabilization} of relative frequencies,
if, for each fixed $\alpha \in L,$ there exists the
limit 
\begin{equation}
\label{l0}
{\bf p} (\alpha)=\lim_{N\to \infty} \nu_N(\alpha) .
\end{equation}
This limit is said to be a probability of $\alpha.$
This probability can be extended to the field of all subsets of $L.$ For each $B\subset L,$ we set
\begin{equation}
\label{l1}
{\bf p} (B)= \lim_{N\to \infty} \nu_N(\alpha \in B)= \lim_{N\to \infty} \sum_{\alpha\in B} \nu_N(\alpha)=
\sum_{\alpha\in B} {\bf p}(\alpha)\;.
\end{equation}

In this paper sequence (\ref{la1}) which satisfies to the principle of the {\it statistical stabilization}
will be called a {\it collective}. We shall not consider so called principle of {\it randomness,} see
[10] for the details. On one hand, randomness could not be defined on the mathematical level of rigorousness
in the von Mises framework. The standard mathematically correct definition of randomness is based on
recursive statistical tests of Martin-L\"of, see, for example, [8].
However, this approach is far from the original frequency framework.
On the other hand, von Mises' principle of randomness is not directly related to
our frequency analysis of EPR-Bell arguments. 
We shall be interested only in the statistical stabilization of relative frequencies.

${\bf p}$ is said to be a {\it probability distribution} of the collective $x.$
We will often use the symbols
${\bf p}(B;x)$ and $\nu_N(B;x), n_N(B;x), B\subset L,$ to indicate dependence on the concrete collective
$x.$
The frequency probability formalism is not a calculus of probabilities. It is
a {\it calculus of collectives.} Instead of operations for probabilities, 
we define operations for collectives.

 An operation of {\it combining of collectives} will play the
crucial role in our analysis of probabilistic foundations of Bell's arguments.
Let $x=(x_j)$ and $y=(y_j)$ be two collectives with label
	sets $L_x$ and $L_y$, respectively. We define a new sequence
$$
z=(z_j),\ z_j=\{ x_j,y_j\}\;.
$$
We remark that in general $z$ is not a collective.
	Let $\alpha \in L_x$ and $\beta\in L_y$. Among the first $N$ elements of
	$z$ there are $n_N(\alpha ;z)$ elements with the first component equal
	to $\alpha $. As $n_N(\alpha ;z)=n_N(\alpha ;x)$ is a number of $x_j=\alpha$ among the
	first $N$ elements of $x$, we obtain that
	$\lim_{N\to\infty}\frac{n_N(\alpha;z)}{N}={\bf p}(\alpha;x)$. Among these
	$n_N(\alpha;z)$ elements, there are a number, say $n_N(\beta/\alpha;z)$ whose
	second component is equal to $\beta$. The frequency $\nu_N(\alpha,\beta;z)$
	of elements of the sequence $z$ labeled $(\alpha,\beta)$ will then be
	$$
	\frac{n_N(\beta/\alpha;z)}{N}=\; \; \frac{n_N(\beta/\alpha;z)}{n_N(\alpha;z)}\; \; \frac{n_N(\alpha;z)}{N}\;.
	$$
	We set $\nu_N(\beta/\alpha;z)=\frac{n_N(\beta/\alpha;z)}{n_N(\alpha;z)}$. Let us assume
	that, for each $\alpha\in L_x$, the subsequence $y(\alpha)$ of $y$ which
	is obtained by choosing $y_j$ such that $x_j=\alpha$ is a
	collective.
	Then, for $\alpha\in L_x$, $\beta\in L_y$, there exists
\begin{equation}
\label{l3}
{\bf p}(\beta/\alpha;z)=\lim_{N\to\infty}\nu_N(\beta/\alpha;z)=\lim_{N\to\infty}\nu_N(\beta;y(\alpha))=
{\bf p}(\beta;y(\alpha)).
\end{equation}
	We have
$
		\sum_{\beta\in L_y}{\bf p}(\beta/\alpha;z)=1. 
$
	The existence of ${\bf p}(\beta/\alpha;z)$ implies the existence of
	${\bf p}(\alpha,\beta;z)=\lim_{N\to\infty}\nu_N(\alpha,\beta;z)$.
	Moreover, we have
\begin{equation}
\label{l2}
{\bf p}(\alpha,\beta;z)={\bf p}(\alpha;x) \; {\bf p}(\beta/\alpha;z)
\end{equation}
	and
$
		{\bf p}(\beta/\alpha;z)={\bf p}(\alpha,\beta;z)/{\bf p}(\alpha;x), 
$
	if ${\bf p}(\alpha;x)\ne 0$.
	
	Thus in this case the sequence $z$ is a collective and  the
	probability distribution ${\bf p}(\alpha,\beta;z)$ is well defined. The
	collective $y$ is said to be {\it combinable} with the collective
	$x$. The relation of combining is a symmetric relation on the set of
	pairs of collectives with strictly positive probability
	distributions, see [8].

	Let $x$ and $y$ be collectives. Suppose that they are combinable. The $y$ is said to be
		{\it independent} from $x$ if all collectives $y(\alpha)$, $\alpha\in L_x$, have the same
		probability distribution which coincides with the probability
	distribution ${\bf p}(\beta;y)$ of $y$. This implies that 
	$$
	{\bf p}(\beta/\alpha;z)=\lim_{N\to\infty}\nu_N(\beta/\alpha;z)=\lim_{N\to\infty}\nu_N(\beta;y(\alpha))
	={\bf p}(\beta;y)\; .
	$$
Here the conditional probability ${\bf p}(\beta/\alpha;z)$ does not depend on $\alpha.$ Hence
	$$
			{\bf p}(\alpha,\beta;z)={\bf p}(\alpha;x)\; {\bf p}(\beta;y),\,\alpha\in L_x,\, \beta\in L_y.
$$

From the physical viewpoint the notion of independent collectives is more natural 
than the notion of independent events in the conventional probability theory in that
the relation ${\bf p}(\alpha,\beta)= {\bf p}(\alpha) {\bf p}(\beta)$ can hold just occasionally 
as the result of a game with numbers, see [10] or [8], p.53.

\section{FREQUENCY ANALYSIS}

 We consider the standard EPR framework. 
Settings of measurement apparatuses 
for particles 1 and 2, respectively,
will be denoted, respectively, by
$a, a^\prime,...$ and $b ,b^\prime, ,... .$ 
In experiments with spin-1/2 particles these setting are given by angles for
axes for measurements of spin projections.
Corresponding physical observables will be denoted by symbols $A, A^\prime,...$
and $B, B^\prime,...$ For simplicity, values of these observables will be denoted below by the
same symbol, e.g.\ $A=\epsilon$, and are supposed to equal $\pm1$.
In the EPR experiments these are measurement outcomes for spin-1/2 particles.

Hidden variables are denoted by $\lambda.$
The most important part of frequency analysis of the EPR-Bell arguments will 
be performed under the assumption that the set of hidden variables is finite, 
$\Lambda=\{1,2,...,M\}.$ 
Such an assumption essentially simplifies frequency analysis and avoids
mathematical technical difficulties.
However, we shall also discuss some frequency effects  which may be induced by
infinite sets of hidden variables. At the end of this section we study the average
procedure with respect to  infinite sets of hidden variables. In appendix 2 
we perform frequency analysis for models with `continuous'
infinite dimensional spaces of hidden variables, spaces of trajectories.

Internal microstates of measurement apparatuses  with settings $a,b,...$
are described by 
variables $\omega_a, \omega_b,....,$ see Bell [1];  sets of these microstates are also finite: 
$\Omega_a = \Omega_b= ...= \{1,..., T \}.$
In fact, this is a contextualistic
model with hidden variables, see Peres in [3] and de Muynck et al. in [4]: 
the value of a physical observable $A$ depends not only on the value
of the hidden variable $\lambda$ for a quantum system, but also on the value of the 
hidden variable $\omega_a$ for  a measurement apparatus  with the setting $a:
A=A(\omega_a,\lambda).$ \footnote{We remark that frequency analysis of the 
EPR-Bell argumentation in the contextualistic framework on the level of physical
observables was performed by Kupczynski [2].}

A sequence of pairs of particles $\pi = \{\pi_j=(\pi^1_j, \pi^2_j), j=1,2,...\}$
is prepared for the same quantum state $\psi.$ By the 
orthodox Copenhagen interpretation $\psi$ gives the complete
description of each quantum system $\pi_j.$
By the statistical interpretation of quantum mechanics, see, for example, [11],
$\psi$ describes  statistical properties of the ensemble $\pi$ of quantum systems, see Peres' book [3]
on an extended discussion.

Let $\lambda_j \in \Lambda, j=1,2,...$ be the value of the hidden variable for the $j$th pair.
For  settings $a$ and $b,$ we consider sequences of pairs
$$
x_{\omega_a, \lambda}= \{ (\omega_{a1}, \lambda_1),...., (\omega_{aN}, \lambda_N),...\}\;,
$$
$$
x_{\omega_b, \lambda}= \{ (\omega_{b1}, \lambda_1),...., (\omega_{bN}, \lambda_N),...\}\;,
$$
where $\omega_{aj}$ and $\omega_{bj}$  are internal states of apparatuses
labeled by $j$
of interactions with particles $\pi^1_j$ and $\pi^2_j,$ respectively.

It should be noticed that there are no physical reasons to suppose that these sequences 
are collectives. Both a preparation device which produces
particles and measurement devices  are complex systems. There are no reasons to suppose that
their micro-fluctuations produce the  statistical stabilization of frequencies:
$$
\nu_N (\omega_a=s, \lambda=k),\; \nu_N (\omega_b=q, \lambda=k),...
$$ 
for fixed $k\in \Lambda,\; s \in \Omega_a, q \in \Omega_b,...$ 
The reader may think that 
the absence of the probability
distributions ${\bf p}(\omega_a=s, \lambda=k), \;  {\bf p} (\omega_b=q, \lambda=k),...,$ 
should contradict 
to the statistical stabilization for the results of observations 
of $A,B,...$ The following considerations show that such a stabilization could 
take place despite fluctuations of frequencies for hidden parameters.

Let us denote by $\Sigma_A(\epsilon)$ the set of 
pairs $(\omega_a, \lambda)$ which produce the value $A=\epsilon$ for the observable $A$. Then
\begin{equation}
\label{q1}
{\bf p}(A=\epsilon) = \lim_{N\to \infty}\sum_{(s,k)\in \Sigma_A(\epsilon)} \nu_N (\omega_a=s, \lambda=k).
\end{equation}
Such a limit of the average with respect to the set $\Sigma_A(\epsilon)$ can exist despite the fluctuations
of frequencies $\nu_N (\omega_a=s, \lambda=k)$ for fixed $s$ and $k,$ see appendix 1.

To continue our analysis, we suppose that, despite the above
critical remarks, sequences $x_{\omega_a, \lambda}$ and $x_{\omega_b, \lambda}$ 
are collectives. Thus, for each setting of a single measurement device, frequency probability
distribution is well defined.
However, in the frequency framework this does not imply that there exists 
frequency probability distribution for each pair of measurement devices.
Therefore we have to study carefully the possibility to combine collectives corresponding to
different measurement devices. Let us write the condition of combining:
$$
\nu_N(\omega_a =s, \lambda = k, \omega_b =q) = \frac{n_N(\omega_a =s, \lambda = k, \omega_b =q)}{N}=
$$
$$
\frac{n_N(\omega_a =s, \lambda = k, \omega_b =q)}{n_N(\omega_a =s, \lambda = k)}\; \; \frac{n_N(\omega_a =s, \lambda = k)}{N}=
$$
$$
\nu_N(\omega_b=q, \lambda=k/\omega_a =s, \lambda = k) \; \nu_N(\omega_a =s, \lambda = k)
\to
$$
$$
{\bf p}(\omega_b=q, \lambda=k/\omega_a =s, \lambda = k) \;{\bf p}(\omega_a =s, \lambda = k)\;,\;  N\to \infty.
$$
Hence, 
$\frac{n_N(\omega_b=q, \lambda=k/\omega_a =s, \lambda = k)}{n_N(\omega_a =s, \lambda = k)}$  
must have the definite limit. 

However, we cannot find physical reasons for such a statistical 
stabilization. Hence, it might be that the probability distribution
${\bf p}(\omega_a =s, \lambda = k, \omega_b =q)$ does not exist, despite the fact that
both probability distributions ${\bf p} (\omega_a =s, \lambda = k)$ and 
${\bf p} (\omega_b=q, \lambda=k)$ are well defined.
The case in that the probabilities
${\bf p} (\omega_a =s, \lambda = k), {\bf p} (\omega_b=q, \lambda=k)$ are well 
defined, but the probability ${\bf p}(\omega_a =s, \lambda = k, \omega_b =q)$ fluctuates can be illustrated
by the following example.

{\bf Example 3.1.} (Uncombinable collectives). {\small Let $D$ be the set of even numbers.
Take any subset $C \subset D$ 
such that 
$$
\frac{1}{N}|C \cap \{1,2,\cdots, N \}|
$$
is oscillating. Here the symbol $\vert O\vert$ denotes the number of elements in
the set $O.$ There happen two cases:
$C \cap \{2n \}=\{ 2n \}$ or $=\emptyset$.
Set
$$
M = C \cup \{2n-1: C \cap \{2n \} = \emptyset\}.
$$
Suppose that, in the sequence $x_{\omega_a, \lambda},$ we have $\omega_a=s$ and $\lambda=k$ for
trails $j\in D,$ and, in the sequence  $x_{\omega_b, \lambda},$ we have $\omega_b=q$ and $\lambda=k$ for
trails $j \in M.$ Both frequency probabilities 
${\bf p} (\omega_a =s, \lambda = k)$ and  ${\bf p} (\omega_b=q, \lambda=k)$ are well defined
and equal to 1/2. However, the probability ${\bf p}(\omega_a =s, \lambda = k, \omega_b =q)$
is not defined.}

To continue our analysis, we suppose that, despite the above
critical remarks, collectives $x_{\omega_a, \lambda}$ and $x_{\omega_b, \lambda}$ 
are combinable. Thus the simultaneous probability distribution 
${\bf p}(\omega_a =s, \lambda = k, \omega_b =q)$ is well defined. To proceed the derivation of Bell-type
inequalities, we have to use the BCHS factorization condition
\begin{equation}
\label{f}
{\bf p}(\omega_a =s, \lambda = k, \omega_b =q) ={\bf p}(\omega_a =s, \lambda = k)\;
{\bf p}(\omega_b =q, \lambda = k)\; .
\end{equation}
This is the condition of independence of collectives.
Hence, to obtain Bell-type inequalities, we have  to
suppose that collectives $x_{\omega_a, \lambda}$ and $x_{\omega_b, \lambda}$ 
are independent. However,
they both contain the same parameter $\lambda.$ This is a kind of constraint.
There must be special physical arguments which would imply that
in the EPR-experiment these collectives are independent despite the $\lambda$-constraint.

Thus our frequency analysis demonstrated that there are at least three probabilistic assumptions 
which are used to obtain Bell-type inequalities in the framework with hidden variables:
1) existence of collectives; 2) possibility of combining; 3) independence. Each 
of these assumptions may be violated for actual quantum processes.

 Typically Bell's framework 
for the EPR experiment is considered without the use of hidden variables for
apparatuses $\omega_a, \omega_b,...$ In such a case only probabilities ${\bf p}(A=\epsilon_1, \lambda=k),
{\bf p}(B=\epsilon_2, \lambda=k), \epsilon_j=\pm 1,$
are used in derivations of Bell-type inequalities. Thus in the frequency analysis we must consider sequences
\begin{equation}
\label{la}
x_{A, \lambda}= \{ (A_1, \lambda_1),...., (A_N, \lambda_N),...\}\; ,
\end{equation}
\begin{equation}
\label{lb}
x_{B, \lambda}= \{ (B_1, \lambda_1),...., (B_N, \lambda_N),...\}\;,
\end{equation}
where $A_j$ and $B_j$  are the $j$th results for observables $A$ and $B.$ 
Here we have similar problems with existence, combining and independence of collectives.
If we even suppose that the sequences $x_{A, \lambda}, x_{B, \lambda},...$ are combinable
collectives, then derivations of Bell-type inequalities will be possible only 
under the assumption that these collectives are independent. Independence of collectives
is equivalent to the factorization of the simultaneous probability distribution:
\begin{equation}
\label{f2}
{\bf p}(A=\epsilon_1,\lambda=k,  B=\epsilon_2; x_{A,\lambda, B})= 
{\bf p}(A=\epsilon_1,\lambda=k; x_{A,\lambda})\; {\bf p}(B=\epsilon_2,\lambda=k; x_{B,\lambda})\;.
\end{equation}
As in the above considerations, independence of these collectives is a rather
doubtful assumption, since both collectives contain the same hidden parameter $\lambda.$

We now discuss the possibility of the transition from
probabilities ${\bf p}(\omega_a=s, \lambda=k)$ to probabilities ${\bf p}(A=\epsilon, \lambda=k).$
\footnote{Such a transition is not so trivial. It was evident even for authors using Kolmogorov's 
measure theoretical viewpoint to probability, see  Shimony [3] and Shimony, Clauser, Horne [2].}

Let $\epsilon=\pm 1, k \in \Lambda.$ Set
$$
\sigma_A(\epsilon;k) =\{ s\in \Omega_a: A(s,k)= \epsilon\},
$$
where $A= A(\omega_a, \lambda)$ is the result of a measurement for the state $\omega_a$ of an apparatus with setting
$a$ and the state $\lambda$ of a quantum particle. Suppose that $x_{\omega_a, \lambda}$ is a collective.
The frequency probabilities ${\bf p}(\omega_a = s, \lambda = k)$ are well defined. We have
$$
{\bf p}(A = \epsilon, \lambda = k; x_{A,\lambda})= \lim_{N\to \infty}
\sum_{s \in \sigma_A(\epsilon;k)} \nu_N(\omega_a=s, \lambda=k;x_{\omega_a, \lambda}).
$$
If the set $\Omega_a$ of microstates of apparatus is finite, then 
we have
\begin{equation}
\label{l7}
\lim_{N\to \infty} \sum_s =\sum_s \lim_{N\to \infty}
\end{equation}
We obtain that the probability ${\bf p}(A=\epsilon, \lambda=k; x_{A,\lambda})$
is well defined. Thus $x_{A,\lambda}$ is a collective. As usual, we have
$$
{\bf p}(A=\epsilon, \lambda=k; x_{A,\lambda})=
\sum_{s \in \sigma_A(\epsilon, k)} {\bf p}(\omega_a=s, \lambda=k;x_{\omega_a, \lambda}).
$$

Suppose that $\Omega_a$ is infinite. Then, in general, we do not have (\ref{l7}).
Thus the assumption that $x_{\omega_a, \lambda}$ is a collective need not imply that 
$x_{A,\lambda}$ is a collective.

{\bf Remark 3.1.} {\small The solution which is proposed in this paper, namely to abandon Kolmogorov probability
theory for von Mises frequency theory, seems to be too easy one, because it does not explain why
Kolmogorov's theory has had so much success in the classical domain, and why this is different 
for quantum mechanics. We can present some speculations on this problem. It might be that 
statistical ensembles which are used in quantum experiments are not sufficiently large to 
produce the statistical stabilization of relative frequencies for hidden variables of quantum systems
and measurement devices. Therefore corresponding frequencies may fluctuate from run to run of 
an experiment. Hence we could not use `constant probabilities', Kolmogorov probabilities. In particular,
we can mention the Bohm-Hiley speculation on complex structures of quantum particles, [12]. Such  complex structures
can be described by spaces of hidden variables of a large cardinality. Different runs of an experiment
may contain quantum particles  with different distributions of hidden parameters.}

{\bf Remark 3.2.} {\small The condition (\ref{f2}) is often interpreted
as the condition of {\it nonlocality.} \footnote{More neutral terms are used by some authors. For example,
A. Shimony called this condition `outcome independence', [3]. De Muynck [4] used the term `conditional statistical
independence.'} Such an interpretation of (\ref{f2})  implies speculations on 
impossibility to use local realism in quantum theory.
However, in the frequency framework (\ref{f2}) has no relation to nonlocality.
One of the reasons for different interpretations of
the violation of factorization
condition (\ref{f2}) is a difference in views to conditional 
probability in the conventional and frequency theories of probabilities.
In the conventional approach  
${\bf p}(U/V) \not= {\bf p}(U)$ implies that the event $U$ depends on the event $V.$
In the EPR framework the violation of (\ref{f2}) implies that the 
event $U = \{$ obtain the value $B= \epsilon_2$ for a particle 2 with $\lambda=k\}$ depends on the event
$V=\{$ obtain the value $A=\epsilon_1$ for a particle 1 with $\lambda=k\}.$
In principle such a dependence of events may be interpreted as an evidence of nonlocalty.
In the frequency framework conditional dependence (or independence) is related not 
to events, but to collectives. Thus the violation of condition (\ref{f}) only implies 
that collectives are dependent.}

We remark that there were numerous discussions on the possibility to use `nonlocality condition'
\begin{equation}
\label{nf}
{\bf p}(A=\epsilon_1,\lambda=k,  B=\epsilon_2)\not = {\bf p}(A=\epsilon_1,\lambda=k)\; {\bf p}(B=\epsilon_2,\lambda=k)
\end{equation}
for the  transmission of information, see, for example, [3]. Typically such a  transmission of information   
was connected with `essentially quantum' properties, so called entanglement.
However, the standard scheme can be applied to transfer information with the aid of any
two dependent collectives which are combinable. 
Let $u=(u_j)$ and $v=(v_j)$ be dependent collectives and let, as usual,
$\epsilon_1, \epsilon_2=\pm 1.$
As they are combinable, conditional probabilities 
$$
{\bf p}(v=\epsilon_2/u=\epsilon_1)=\lim_{N\to \infty} \nu_N(v=\epsilon_2;v(\epsilon_1))
$$
are well defined. Here, as usual, $v(\epsilon_1)$ is a collective obtained from $v$ by the choice
of subsequence $v_{j_k}$ such that $u_{j_k}=\epsilon_1.$
As collectives are dependent, we have, for example,
$$
{\bf p}_1= {\bf p}(v=1/u=+1) \not= {\bf p}_2= {\bf p}(v= 1/u=-1).
$$
We can proceed in the same way as in all `quantum stories'. Bob prepares a statistical ensemble
of pairs which components are described by collectives $u$ and $v$ respectively. He chooses
subcollective $v(+1)$ and sends it to Alice. If Alice knows the relation between probabilities,
she can easily rediscover the bit of information. 

\section{LINKS TO SOME MEASURE-THEORETICAL RESULTS}

In this section we present connections with some well known results on Bell's 
inequality which were obtained on the basis of Kolmogorov probability model.
It was proved by Fine and Rastall [6] that Bell's inequality is equivalent 
to the existence of the simultaneous probability distribution for physical 
observables $A, A^\prime, B$ corresponding to three different
settings $a, a^\prime, b$ of measurement apparatuses.
As usual in this paper symbols $a, a^\prime$  and $b$ are used, respectively,
for settings of measurement devices for the first particle and second particle.  We analyse
the Fine-Rastall framework from the frequency viewpoint. 

As it has been mentioned, in the frequency theory we could not consider a probability
distribution without relation to some collective. However, the object which is
called a `probability distribution' in the Fine-Rastall framework has no relation 
to a collective. So such an object has no probabilistic and, consequently, physical
meaning from the frequency viewpoint.\footnote{Eberhard [2] rightly pointed out that
Fine's statements contain rather unclear words on simultaneous probability distribution: ``well defined."}
It seems that the Fine-Rastall condition is just a purely mathematical constraint.

If we accept the use of counterfactuals, see Peres [2] and [3] on an extended discussion,
then we can continue frequency analysis 
of the  Fine-Rastall arguments. Beside of collectives $x_{A,\lambda}= \{ (A_j, \lambda_j), j=1,2,...\},
x_{B,\lambda}= \{ (B_j, \lambda_j), j=1,2,...\},$ we can consider  `gedanken kollektiv'
$x_{A^\prime,\lambda}= \{ (A^\prime_j, \lambda_j), j=1,2,...\}.$ Suppose that three collectives
are combinable. There exists the simultaneous probability distribution $(\epsilon_1, \epsilon_2, \epsilon_3= \pm 1):$
$$
{\bf p} (A=\epsilon_1, B=\epsilon_2, A^\prime= \epsilon_3, \lambda=k)
$$
$$
= \lim_{N\to \infty}
\frac{1}{N} \nu_N(A=\epsilon_1, B=\epsilon_2, A^\prime= \epsilon_3, \lambda=k; x_{A,B,A^\prime,\lambda})\;.
$$
The average with respect to $\lambda$ (if such a procedure is justified) gives  the 
simultaneous probability distribution: 
\begin{equation}
\label{lq}
{\bf p} (A=\epsilon_1, B=\epsilon_2, A^\prime = \epsilon_3)=\lim_{N\to \infty}
\frac{1}{N} \nu_N(A=\epsilon_1, B=\epsilon_2, A^\prime= \epsilon_3; x_{A B A^\prime}) \;.
\end{equation}
In this case we can apply the Fine-Rastall theory and obtain Bell's inequality
without the assumption that collectives $x_{A,\lambda}$ and $x_{B,\lambda}$ are independent, i.e.,
without factorization condition (\ref{f2}). 

We now suppose that three  collectives $x_{A,\lambda}, x_{B,\lambda}, x_{A^\prime,\lambda}$ 
are not combinable. Thus limit (\ref{lq}) does not exist. There is no simultaneous probability distribution
${\bf p} (A=\epsilon_1, B=\epsilon_2, A^\prime= \epsilon_3).$ 
However, it can occur that there exists real numbers ${\bf p}_{\epsilon_1\epsilon_2\epsilon_3}\geq 0,
\sum {\bf p}_{\epsilon_1\epsilon_2\epsilon_3} =1$ such that ${\bf p}(a= \epsilon_1, b= \epsilon_2; x_{ab})=
\sum_{\epsilon_3} {\bf p}_{\epsilon_1\epsilon_2\epsilon_3}.$ By the Fine-Rastall result we have Bell's inequality.

This identification of mathematical Fine-Rastall constants with physical probabilities is the root 
of some misunderstanding of the role of the Fine-Rastall result. This result is often interpreted
as the demonstration that BCHS locality condition is not directly related
to Bell's inequality. The violation of Bell's inequality is connected with the fact that
observables $A$ and $A^\prime$ are incompatible. This implies the absence of the simultaneous probability 
distribution even for two observables $A$ and $A^\prime.$
However, such an inference might be only done if
we could prove that Bell's inequality must imply the existence of frequency
probability distribution (\ref{lq}). However, it seems to be impossible to obtain such a result.

\bigskip

{\bf Conclusion}
{\it In the frequency approach (if we follow to R. von Mises and define
probabilities as limits of relative frequencies and not as abstract Kolmogorov measures) 
arguments related to locality and  determinism  do not play an important role
in Bell's framework. 

In this approach formal probabilities $p(a=\pm 1, b=\pm 1/ \lambda)$ which are 
used by many authors need not exist at all. It is a rather normal situation in the frequency approach. 
Moreover, here the BCHS locality condition does not have the standard locality interpretation. It was  rightly
called "outcome independence condition" [3]. However, everybody who works in Kolmogorov's
axiomatic approach, conventional probability theory, considers dependence or independence
as  dependence or independence of EVENTS. Of course, such a viewpoint implies nonlocality:
one event depends on another. In von Mises' approach dependence or independence
has the meaning of dependence or independence of collectives, random sequences.
Such a dependence is a consequence of the simultaneous preparation procedure
for two collectives. Of course, this does not exclude the possibility that some 
nonlocal effects also play some role in the creation of such a dependence.}

\medskip

\medskip

{\bf APPENDIX 1.}

\medskip

{\small Let us consider motion of a particle on the line. A preparation procedure $\Pi$
produces particles with velocities $v=+1$ and $v=-1.$ Suppose that $\Pi$ cannot control 
(even statistically) proportion of particles moving in positive and negative directions.
This proportion fluctuates from run to run. Mathematically we can describe this situation
as the absence of the statistical stabilization in the sequence:
$x_v = (v_1, v_2,..., v_N,...), \; v_j= \pm 1\;,$ of velocities of particles. For example,
let relative frequencies 
$
\nu_N(v=+1)\approx  \sin^2 \phi_N$ and 
$\nu_N(v= -1)\approx \cos^2 \phi_N.$
If `phases' $\phi_N$ do not stabilize $(\rm{mod}\; 2 \pi)$ when 
$N \to \infty,$ then frequencies $\nu_N(v= +1), \nu_N(v= -1)$ fluctuate when $N\to \infty.$
Hence the sequence $x_{v}$ is not a collective. Thus the principle of the statistical
stabilization is violated. Suppose that we have an apparatus to measure the energy of a
particle: $E= v^2/2.$ We obtain that $E= 1/2$  with the probability one.
Suppose that we cannot measure the velocity.  Then we would not know that the
measured value $E=1/2$ is produced by chaotic fluctuations of the (objective) velocity.

A slight modification can give an example in that `fluctuating microreality'
produces states which are not eigenstates of the $E.$ Let $v=\pm 1, \pm 1/2$ and let
$
\nu_N(v=+1)= \nu_N(v=-1/2) \approx \frac{1}{2} \sin^2 \phi_N$ and 
$\nu_N(v= -1)=\nu_N(v= +1/2) \approx \frac{1}{2} \cos^2 \phi_N.$
Suppose that again `phases' $\phi_N$ do not stabilize. Thus probabilities
${\bf p}(v= +1), {\bf p}(v= -1), {\bf p}(v= 1/2), {\bf p}(v= -1/2)$ do not exist.
However, the frequency probabilities ${\bf p} (E=1/2), {\bf p} (E= 1/8)$ 
are well defined and equal to 1/2. Suppose that we can measure only
the energy (and cannot observe this oscillation of probabilities 
for the velocity). Then we can, in principle, suppose that there exists
the probability distribution of the velocity in this experiment and use such a distribution
in some considerations. It may be that we do such an illegal trick in Bell's framework.}

\medskip

\medskip

{\bf APPENDIX 2: FREQUENCY ANALYSIS OF TIME-AVERAGE MODEL
FOR THE EPR EXPERIMENT}

\medskip

In section 3 we considered a simplified model with finite sets of hidden variables. That model
was useful to find implicit probabilistic assumptions which were used to prove Bell-type inequalities.
However, real processes of measurements could not be described by finite sets of hidden variables.
Processes of  measurements are not $\delta$-function processes. The values of physical 
observables are time averages of hidden variables $\lambda$ and $\omega_a, \omega_b,... $ 
which evolve with time.
In fact, 
$A=A(\xi_a, \eta_a)$ is a functional of trajectories of the microstates of
the apparatus $\xi_a= \omega_a(\cdot)$ and a quantum particle $\eta_a = \lambda_a(\cdot).$ 
There are the initial
conditions $\omega_a(0)=\omega_a^0$ and $\lambda_a(0)= \lambda^0.$
Here $\omega_a^0$ is the microstate of $a$ and $\lambda^0$ is the value of hidden variable
for a quantum particle before interaction.
In general we cannot assume  that
trajectories $\xi_a$ and $\eta_a$ evolve independently.
The interaction between a particle and an
apparatus induces the simultaneous evolution  of $\xi_a$ and $\eta_a.$

Let us consider a series of experiments with correlated particles. For the apparatus $a,$
we have a series of two dimensional trajectories:
\begin{equation}
\label{p1}
x_{u_a} = (u_{a1}, u_{a2},...., u_{aN},....), \; u_j =(\xi_{aj}, \eta_{aj}),
\end{equation}
where $u_{aj}(t)=(\omega_{aj}(t), \lambda_{aj}(t))$ is a solution of the equation:
$$
\frac{d u_{aj}}{d t}= {\cal A}_j(u_{aj}(t)),\; \; u(0)= (\omega_a^0,\lambda^0) \; .
$$
In general the operator of evolution ${\cal A}$ depends on the trial $j$ (uncontrolled fluctuations
of fields), ${\cal A}={\cal A}_j.$ 
The corresponding series of two dimensional trajectories for the apparatus $b$ is denoted by the symbol
$x_{u_b}.$

We again consider the problem of the existence of collectives.
Here we have to be more careful with the choice of a label set. Suppose that all trajectories
are continuous. Denote by the symbol $C$ the space of continuous trajectories endowed 
with the uniform norm.
Denote by symbol ${\cal B}(C)$ the $\sigma$-field of Borel subsets of the metric space $C.$
In principle, we are interested in the statistical stabilization of frequencies
$\nu_N(u \in D \times E; x_{u_a})= n_N(u \in D \times E; x_{u_a})/N,$ where sets $D, E \in {\cal B}(C).$
It is well known [10] that in general 
there is no such a stabilization for all Borel sets even in the finite dimensional case. 
Thus sequence (\ref{p1}) need not be a collective with respect to the 
set of labels 
$$
L= \{ D \times E: D, E \in {\cal B}(C)\}.
$$ 
The existence of the Kolmogorov probability
distribution ${\bf p}(\xi_a \in D, \eta_a \in E)$ 
on the set of hidden parameters $(\xi_a, \eta_a)$ is an additional mathematical assumption.

To continue our analysis, we suppose that $x_{u_a}$ is a collective with respect to some
subfield ${\cal B}_0(C)$ of ${\cal B}(C).$ Thus
$$
{\bf p}(\xi_a \in D, \eta_a \in E)= \lim_{N\to \infty}
\nu_N(u \in D \times E; x_{u_a}), \; \; D,E \in {\cal B}_0(C)\; .
$$
Here  the label set 
$$
L_0
=\{ D \times E: D, E \in {\cal B}_0(C) \}.
$$ 
In general ${\bf p}$ is not a Kolmogorov $\sigma$-additive
measure, but only a finite additive measure. Standard derivations of Bell-type 
inequalities are blocked by the purely mathematical problem: integration with respect 
to finite-additive measures.

To continue our analysis, we suppose that we could solve mathematical problems related to integration
with respect to finite additive measures.
However, the derivation would be again blocked, because collectives
$x_{\xi_a}$ and $x_{\eta_a}$ consisting of  trajectories $\omega_a(t)$ and $\lambda_a(t),$ respectively,
are not independent. 
Dependence is generated in the process of evolution via the mixing by the evolution operator ${\cal A}.$
 There is no factorization condition:
$
{\bf p}(\xi_a \in D, \eta_a \in E; x_{u_a})
={\bf p}(\xi_a \in D; x_{\xi_a}) \; {\bf p}(\eta_a \in E; x_{\eta_a})
$
even for $D, E \in {\cal B}_0(C).$

Despite all of these problems we continue our analysis.
In principle collectives
 may be not  combinable even 
with respect to the label set $L_0 \times L_0.$ 
Nevertheless, suppose that in the EPR-Bell framework
they are combinable. 
Hence there exists
a finite additive measure
${\bf p} (\xi_a \in D_1, \eta_a \in E_1, \xi_b \in D_2, \eta_b \in E_2).$
Of course, the absence of $\sigma$-additivity is a mathematical problem. However, the main
problem is that collectives $x_{u_a}$ and $x_{u_b}$ are not independent, because trajectories
$u_a$ and $u_b$ are connected at the initial instant of time by the constraint:
$\lambda_a(0)=\lambda_b(0)=\lambda^0.$ 

In the present model collectives
corresponding to different measurement apparatuses are always dependent.
There is no factorization 
$$
{\bf p} (\xi_a \in D_1, \eta_a \in E_1, \xi_b \in D_2, \eta_b \in E_2)=
{\bf p} (\xi_a \in D_1, \eta_a \in E_1)\; {\bf p} (\xi_b \in D_2, \eta_b \in E_2)\;.
$$
In general there is no Bell's  inequality. 

\bigskip

\bigskip

{\bf REFERENCES}

1. J. S. Bell,  {\it Rev. Mod. Phys.,} {\bf 38}, 447 (1966).
J. S. Bell, {\it Speakable and unspeakable in quantum mechanics}
(Cambridge Univ. Press, Cambridge, 1987).

2. J. F. Clauser, M. A. Horne, A. Shimony and R. A. Holt,
{\it Phys. Rev. Letters,} {\bf 23}, 880 (1969);
J. F. Clauser  and   A. Shimony,  {\it Rep. Progr. Phys.,}
{\bf 41}, 1881 (1978);
 A. Aspect,  J. Dalibard  and   G. Roger, 
{\it Phys. Rev. Lett.,} {\bf 49}, 1804 (1982);
 A. Home  and   F. Selleri, {\it Riv. Nuovo Cimento,} {\bf 14},
1 (1991); H. P. Stapp, {\it Phys. Rev.,} {\bf A  3}, 1303 (1971);
P. H. Eberhard, {\it Il Nuovo Cimento,} {\bf B 38}, 75 (1977); {\it Phys. Rev. Lett.,}
{\bf 49}, 1474 (1982);
A. Peres, {\it Am. J. of Physics,} {\bf 46}, 745 (1978); 
{\it Found. Phys.,} {\bf 14}, 1131 (1984); {\bf 16}, 573 (1986);
P. H. Eberhard,  {\it Il Nuovo Cimento,} {\bf B 46}, 392 (1978); J. Jarrett, {\it Nous,} {\bf 18},
569 (1984); M. Kupczynski, {\it Phys. Lett.,} {\bf A 121}, 205 (1987).

3. B. d'Espagnat, {\it Veiled Reality. An Analysis of Present-Day
Quantum Mechanical Concepts} (Addison-Wesley, Reading, MA, 1995);
A. Shimony, {\it Search for a naturalistic world view} (Cambridge Univ. Press, Cambridge, 1993);
A. Peres, {\it Quantum Theory: Concepts and Methods} (Kluwer Academic, Dordrecht, 1994).

4. L. de Broglie, {\it La thermodynamique de la particule isolee} (Gauthier-Villars, Paris,
1964); G. Lochak, {\it Found. Phys.,} {\bf 6}, 173 (1976);
E. Nelson, {\it Quantum fluctuation} (Princeton Univ. Press, Princeton, 1985);
W. de Muynck and W. De Baere W.,
{\it Ann. Israel Phys. Soc.,} {\bf 12}, 1 (1996);
W. de Muynck, W. De Baere  and  H. Martens,
{\it Found. Phys.,} {\bf 24}, 1589 (1994);
W. de Muynck  and  J. T. van Stekelenborg,  {\it Annalen der Physik,} {\bf 45},
222 (1988).

5. L. Accardi, {\it Urne e Camaleoni: Dialogo sulla realta,
le leggi del caso e la teoria quantistica} (Il Saggiatore, Rome, 1997);
Accardi  L., The probabilistic roots of the quantum mechanical paradoxes.
{\it The wave--particle dualism. A tribute to Louis de Broglie on his 90th 
Birthday}, Edited by S. Diner, D. Fargue,  G. Lochak and F. Selleri, 47--55
(D. Reidel Publ. Company, Dordrecht,1970).

6. I. Pitowsky,  {\it Phys. Rev. Lett,} {\bf 48},  1299 (1982);
{\it Phys. Rev.} {\bf A 27}, 2316 (1983);
S.P. Gudder,  {\it J. Math Phys.,} {\bf 25}, 2397 (1984);
A. Fine,  {\it Phys. Rev. Lett.,} {\bf 48}, 291 (1982);
P. Rastal, {\it  Found. Phys.,} {\bf 13}, 555 (1983);
W. Muckenheim,  {\it Phys. Rep.,} {\bf 133}, 338 (1986);
W. De Baere,  {\it Lett. Nuovo Cimento,} {\bf 39}, 234 (1984);
{\bf 40}, 448 (1984).

7. A. Yu. Khrennikov,  {\it Dokl. Akad. Nauk SSSR,} ser. Matem.,
{\bf 322},  1075 (1992); {\it J. Math. Phys.,} {\bf 32}, 932 (1991);
{\it Phys. Lett.,} {\bf A 200}, 119 (1995);
{\it Physica,} {\bf A 215}, 577 (1995);   {\it Int. J. Theor. Phys.,} {\bf 34},
2423 (1995);  {\it J. Math. Phys.,} {\bf 36},
6625 (1995);
A.Yu. Khrennikov, {\it $p$-adic valued distributions in 
mathematical physics} (Kluwer Academic, Dordrecht, 1994);
A.Yu. Khrennikov, {\it Non-Archimedean analysis: quantum
paradoxes, dynamical systems and biological models}
(Kluwer Academic, Dordrecht, 1997).

8. A. Yu. Khrennikov, {\it Bell and Kolmogorov: probability, reality and nonlocality.}
Reports of V\"axj\"o Univ., N. 13 (1999); A. Yu. Khrennikov, {\it Interpretations of probability}
(VSP Int. Sc. Publ., Utrecht, 1999).

9. A. N. Kolmogorov, {\it Grundbegriffe der Wahrscheinlichkeitsrechnung}
(Springer Verlag, Berlin, 1933); reprinted:
{\it Foundations of the Probability Theory}
(Chelsea Publ. Comp., New York, 1956).

10. R.  von Mises, {\it The mathematical theory of probability and
 statistics} (Academic, London, 1964).

11. L. E. Ballentine,  {\it Rev. Mod. Phys.,} {\bf 42}, 358 (1970). 

12.   D. Bohm and B. Hiley, {\it The undivided universe:
an ontological interpretation of quantum mechanics} (Routledge and Kegan Paul, 
London, 1993).

\end{document}